# Inventory record inaccuracy in grocery retailing: Impact of promotions and product perishability, and targeted effect of audits

Yacine Rekik

emlyon business school, France

yrekik@em-lyon.com

Rogelio Oliva

Mays Business School Texas A&M University College Station, TX, USA

roliva@tamu.edu

Aris Syntetos

Cardiff Business School Cardiff University Cardiff, Wales, UK

syntetosa@cardiff.ac.uk

Christoph Glock*

Institute of Production and Supply Chain Management Technical University of Darmstadt Darmstadt, Hessen, Germany

glock@pscm.tu-darmstadt.de

# Inventory record inaccuracy in grocery retailing: Impact of promotions and product perishability, and targeted effect of audits

## Abstract

We report the results of a study to identify and quantify drivers of inventory record inaccuracy (IRI) in a grocery retailing environment, a context characterized by frequent promotions and a high share of perishable items. The analysis covers ~24,000 stock keeping units (SKUs) sold in 11 stores. We find that IRI is positively associated with average inventory level, restocking frequency, and whether the item is perishable, and negatively associated with promotional activity. We also conduct a field quasi-experiment to assess the marginal effect of stockcounts on sales. While performing an inventory audit is found to lead to an 11% store-wide sales lift, the audit has heterogeneous effects with all the sales lift concentrated on items exhibiting negative IRI (i.e., where system inventory is greater than actual inventory). The benefits of inventory audits are also found to be more pronounced on perishable items, that are associated with higher IRI levels. Our findings inform retailers on the appropriate allocation of effort to improve IRI and reframe stock counting as a sales-increasing strategy rather than a cost-intensive necessity.

## 1. Introduction

To manage inventory levels, companies usually employ software tools that use inventory and sales data to automatically forecast demand and replenish warehouses and store shelves. The global retail analytics market attained a size of around $7.4 billion in 2022 and is expected to reach $40.4 billion by 2032, growing at a compound annual growth rate of 18.6% during the forecast period (DataHorizon Research, 2023). The analytics market consists of both software and associated services. The key analytical software for supply chains within the retail sector is built with a focus on inventory management, supply and demand forecasting, and vendor management. Companies' investments in inventory software are immense.

One weakness of automatic inventory management systems though is that replenishments are triggered based on inventory levels recorded in the system, which can differ from the inventory available in the retail store or warehouse. Discrepancies between the actual and the recorded inventory level are termed Inventory Record Inaccuracies (IRI; e.g., Chuang et al., 2016; Goyal et al., 2016), and they have been reported to greatly impact the retail sector; earlier research has found that between 20% and 70% of inventory records are incorrect (e.g., Millet, 1994; Kang & Gershwin, 2005; DeHoratius & Raman, 2008).

If the inventory management system displays a stock level that is higher than the inventory level available (negative IRI), then inventory is replenished too late, which can lead to stockouts and the company not being able to satisfy customer demand (DeHoratius & Raman, 2008; Rekik et al., 2019). In the opposite case, where the available stock exceeds the recorded one (positive IRI), the system triggers unnecessary replenishments and extra supplies leading to unnecessarily high inventory carrying cost (Hardgrave et al., 2009; Chuang et al., 2016). Regardless of whether inventory discrepancies are positive or negative, inaccurate inventory records prevent the (costly to install and operate) stock management system to unfold its full potential.

Given the detrimental impact of IRI, it is not surprising that there is a well-established research tradition to investigate the causes and consequences of IRI as well as approaches for reducing it (see §2 for an overview). These prior studies, however, were done on IRI in general-merchandise stores, warehouses, and distribution centers (e.g., DeHoratius & Raman, 2008; Kang & Gershwin, 2005; Chuang & Oliva, 2015). Grocery retailing is a context in which two operational features are especially salient: a notable share of stock keeping units (SKUs) is perishable, and products are often subject to promotions. Product perishability introduces additional record-keeping complexity, as SKUs are replenished more frequently and are subject to a set of mandatory shelf-management stacks – including First-In-First-Out (FIFO) rotation, expiration date checks, markdown executing and removal of expired goods (discussed in detail in §3) (Blattberg et al., 1995; van Donselaar et al., 2006; Ailawadi et al., 2009; Akkas et al., 2019). These additional transactions and stricter processes introduce further opportunities for record errors. Promotional activities also come with a set of challenges, as they simultaneously increase sales volatility and demand greater inventory control (e.g., additional monitoring to secure proper reimbursement or processing of discounts). Promotions reallocate managerial effort toward price changes, displays, and compliance checks, altering both the generation and correction mechanisms of IRI (Blattberg et al., 1995; Ailawadi et al., 2009). These conditions make grocery retailing a meaningful context in which to explore how these elements interact to affect IRI.

To theoretically ground these dynamics, we draw on the Attention-Based View (ABV; see Ocasio, 1997), which conceptualizes organizational attention as a scarce and selectively allocated resource that shapes which operational issues are noticed and acted upon. In grocery retailing, inventory management requires continuous attentional engagement across a large number of SKUs and operational touchpoints, making inventory record accuracy critically dependent on how organizational attention is distributed in day-to-day execution. From an ABV perspective, factors such as inventory levels, replenishment frequency, perishability, and promotions shape the salience and attentional demands associated with inventory management tasks, thereby influencing the likelihood that inventory record inaccuracies emerge or remain uncorrected. By applying ABV to store-level inventory management, we extend the theory into an operational setting where attentional demands are distributed across routine execution rather than concentrated in strategic decision-making.

Remarkably, to date there are no empirical studies that explore the drivers and impact of IRI in grocery retailing. This is surprising given the economic importance of the retail sector. In the U.S. alone, retail sales exceeded $7 trillion in 2024, with offline grocery retailing being responsible for about 23% of this value (U.S. Census Bureau, 2025). Reports estimate that fresh departments, which are a sub-category of perishable items, account for more than 40% of grocery retail sales (FoodNavigator, 2024). Ensuring accurate inventory records and effective replenishment processes is hence particularly critical for this product category. Against this background, the first study reported in this paper assesses the impact of perishability (and the associated effects it has on inventory levels and replenishment frequency) and promotions on IRI. Our study thus contributes new construct-level drivers relevant to the retail sector (perishability; promotion-period oversight).

Although it is clear that inaccurate inventory records hamper the functionality of automated restocking systems, little is known about the magnitude of the impact of IRI on sales revenue. Earlier research has used simulation for estimating the impact of "inventory freezing"[1] on sales (e.g., DeHoratius et al., 2008) or analyzed the sales improvement potential of stockcounts for a reduced set of SKUs (e.g., Chuang et al., 2016). Given the limited scope of earlier research on the sales impact of IRI, the second study presented in this paper reports the results of a field quasi-experiment to evaluate how a storewide audit (as opposed to counts of individual SKUs) impacts sales. We also use the quasi-experimental context to assess the temporal impact of audits and, controlling for product attributes and inventory states, we are able to use the treatment results to identify areas where the audits are more impactful.

Our studies were undertaken in collaboration with a major European grocery retailer. In the first study, we find that, after controlling for the major SKU level drivers identified by earlier research, a SKU's IRI is positively associated with its average inventory level, its restocking frequency, and whether the item is perishable, and negatively associated with its promotional activity. The second study finds that performing audits to correct such inaccuracies leads to approximately 11% of increased store-wide sales in the first two months after the stock count. More interestingly, we find that, controlling for sales levels and other factors, the stockcount has heterogeneous effects on SKUs with all the sales lift concentrated on items exhibiting negative IRI and that perishable items, that are more propense to exhibit IRI, benefit more from those audits than non-perishable items. We believe these findings will be of great value to retailers to inform their decisions on the appropriate levels of investment that they should put against improving inventory record accuracy, and to discuss stock counting as a sales-increasing strategy rather than a cost-intensive necessity.

[1] "Inventory freezing" refers to an Out-Of-Stock (OOS) item that is not replaced because the information system believes there is inventory available (Kang & Gershwin, 2005).

The remainder of our paper is structured as follows. In §2, we review the relevant literature that informs our study both in terms of the methodological approach we employ and the formation of the questions we seek to answer. The theoretical foundation of our work is discussed in §3, where the hypotheses are developed, and the measures of our analysis established. §4 is devoted to the analysis of the IRI drivers. In §5, we assess the store-wide impact of audits on sales. Theoretical, empirical and managerial insights, and limitations along with an agenda for future relevant research, are considered in §6.

## 2. Literature review

One of the most important objectives in retailing is to ensure high on-shelf availability of products. If customers are after an item that is not available for sale, they may decide to buy it elsewhere, or they may abandon their shopping trip altogether, negatively affecting the sales of other items as well. If stockouts occur frequently, store reputation may suffer, and customers may switch to another brand or store, jeopardizing future sales (Corsten & Gruen, 2003; Ehrenthal & Stölzle, 2010; Goyal et al., 2016). This is particularly problematic in the grocery retail context. Grocery shopping has been described as a habitual process, with customers often following routines in terms of shopping hours, store visits and products purchased (East et al., 1994; Kim & Park, 1997). Routine shoppers face high costs if they switch stores, so they tend to leave out-of-stock items out or buy alternative items. If stockouts occur frequently or for an extended period, then customers may switch stores as well (Campo et al., 2004). As previously discussed, one cause of stockouts is inaccurate inventory records, although the transition is not necessarily linear nor obvious (Chuang et al., 2022). In the case of negative IRI where the inventory system overestimates the available stock, the system orders too late, which may result in a stockout the system is not even aware of (Ehrenthal & Stölzle, 2010).

Inaccurate inventory records have attracted quite some attention over the years, with researchers trying to obtain insights both into the sources, the extent and consequences of IRI as well as the effectiveness of possible countermeasures across different sectors. Table A1 in the Appendix summarizes related research with some key insights from these works discussed in the following. This table categorizes studies into three primary domains: the drivers of IRI, strategies for reducing IRI, and the impact of IRI on sales. For each study, we present the main objective, research context, key data attributes, and principal findings.

An early work studying drivers of IRI is the one of Rinehart (1960), who found that around 80% of the discrepancies were caused by the discrepancy correction procedures themselves. DeHoratius and Raman (2008) analyzed stockcount data collected at 37 stores of a major US retailer. The most important IRI drivers they identified are an item's annual selling quantity, its cost, and the frequency of stock audits. The authors also investigated the extent of IRI and found that 65% of the inventory records they analyzed were inaccurate. The absolute difference between the stock on shelf and the inventory record was five units on average per SKU, or 35% of the average number of units found on the shelf for the SKU. Further evidence on the extent of IRI can be found in Kang and Gershwin (2005) and Ishfaq and Raja (2020a, 2020b), who reported IRI rates ranging between 20% and 70%. IRI has also been investigated in contexts other than retail stores, such as in manufacturing warehouses (e.g., Meyer, 1990; Sheppard & Brown, 1993) or retail distribution centers (e.g., Kull et al., 2013; Barratt et al., 2018). Although IRI has been found to be prevalent in these contexts too, its behavior is likely different due to the absence of a major driver of inventory discrepancies: the customer.

In addition to identifying the sources and the extent of IRI, researchers have also investigated the effectiveness of different measures in reducing IRI. Several papers (e.g., Hardgrave et al., 2009; 2013; Bertolini et al., 2015; Goyal et al., 2016) studied how Radio Frequency Identification (RFID) technology can contribute to lowering IRI. These studies showed that using RFID during stockcounts or for tracking stock movements can lead to a substantial decrease in IRI, but that it cannot alleviate the problem completely. Chuang and Oliva (2015) investigated the influence of store staffing levels and operational performance on IRI. In an empirical analysis of longitudinal data from five stores of a global retail chain, they found evidence that full-time labor reduces IRI, whereas part-time labor fails to alleviate it. DeHoratius et al. (2023) evaluated different methods for generating a list of items to be counted in the store to correct IRI. The authors identified a set of relatively simple rule-based approaches that can easily be deployed in practice.

While several researchers have investigated how retail stockouts may affect sales (see Gruen et al., 2002, for an overview), only a few works have addressed the impact of IRI on sales. Ton and Raman (2010) studied how so-called phantom products, i.e., products that are physically present in the store, but only in areas where customers cannot find them, influence sales. The authors found that a one standard deviation increase in the percentage of phantom products led to a 1% decrease in store sales. Phantom products are not necessarily connected to IRI; that is, the inventory records in the system could be accurate, but some of these items – the phantom products – could be misplaced and not accessible to customers. In the context of our analysis, the fraction of phantom products not appropriately recorded in the stock management system would be regarded as positive inventory discrepancy. DeHoratius and Raman (2008) used empirical data to conduct a simulation study on the influence of "*frozen products*" on sales. They estimated that inventory freezing can result in a loss of sales of about 1%. The authors hypothesized that their estimate is only a lower bound on the true sales loss and called for further empirical research in this area. The work that probably comes closest to our investigation is the one of Chuang et al. (2016), who developed a model for predicting out-of-shelf events based on consecutive zero sales-situations. In a field experiment, they used their model to trigger audits of eight different SKUs that suffered from inventory freezing and found that the intervention had a positive effect on sales.

Our investigation departs from, and contributes to, the extant literature in three ways. We map each of these departures onto elements of the typology developed by Makadok et al. (2018) that sets out the ways in which research contributes to theory. First, we evaluate drivers of IRI in a grocery retail environment, i.e., in the presence of perishable products and product promotions. Earlier research has not yet investigated how product perishability and promotions interact with IRI. With respect to our theoretical contribution here, we introduce a new construct and variables as the focal phenomenon. Secondly, we evaluate, through an intervention study, the impact of a store-wide stockcount on sales. The store-wide assessment of the stockcount allows us to empirically assess heterogenous effects of the audit on products with different attributes and different accuracy status. In so doing, theoretically, we introduce a new level of analysis for the effects of audits. Finally, through a dynamic evaluation of the effects of the intervention, we trace the evolution of benefit post stockcount. In developing the scholarship in this way, our work offers the possibility for future research into the appropriate timing of interventions to correct the problem, and as such the opportunity to make new theoretical contributions.

## 3. Empirical and theoretical background

### *3.1. Research setting and data collection*

We worked together with a major European grocery retailer. The partner organization is a higher end supermarket, known for the quality of their products and customer service. They operate stores throughout the UK and in 2024 had an annual sales volume of more than £7 billion. In addition to regular groceries, they sell specialty products, including a wide range of own brand products, and most stores have a bakery, cheese, fish, and meat counters. They also sell household goods, gifts, and clothes. With more than 300 shops in the UK, the company is one of the largest retailers of groceries in the country, also exporting products to many other countries, in some of which they have physical presence.

The retailer uses a stock management system to replenish its stores. The system automatically triggers orders, but store managers have the option to manually adjust or trigger orders if required. In terms of stock audits, the company employs a third party to conduct periodic stockcounts at the store level, up to four times a year. The timing of these stockcounts varies across stores and is typically driven by operational metrics such as self-service penetration, shrinkage and theft exposure, and stock turnover. This standard process applies to the first stage of our analysis, where we relied on historical stockcount data collected independently of our research. However, in the second stage of the study – the quasi-experiment involving a store-wide audit – the timing of the stockcount was scheduled to align with our research design and was not triggered by the usual operational metrics.

Prior to the actual stockcounts, a preparation team provided by the third party prepares the stores for stocktaking by tidying both the backroom and the shop floor (in this step, easy-to-spot misplaced items are identified and brought back to their correct locations). The actual stockcounts then start in the backroom and continue on the shop floor in slow selling areas just before the store closes. The remaining parts of the store are counted outside of business hours.

Although we did not have access to complete statistics for the retailer's full network of over 300 stores, our sample was constructed in close collaboration with the retailer, who ensured that selected stores represented a range of typical operational formats (urban, city center, and suburban), sizes, and service features. This approach increases the practical relevance of our findings but limits the ability to formally compare sample characteristics to the entire network, which we acknowledge as a limitation of this study. A total of 11 stores were selected as representative of the different store types operated by them. The 11 stores under study have different capabilities (ecommerce, self-checkout service, car parking) and different sizes (ranging from 6,663 to 32,032 square feet), different location types (urban, suburban, city center, rural town) and different numbers of suppliers (ranging from 55 to 351 suppliers). A majority of those stores are core type outlets selling general grocery merchandise and some of them are convenience stores focusing on a limited range of everyday products. All stores were ultimately randomly chosen, except for stores 1 and 2 where we involved the retailer in the matching between the stores. This pair of stores was used in the field quasi-experiment to assess the benefit of the stockcounts and the effects of IRI on sales described in §5. Details of the selected stores are presented in Table 1.

**Table 1: Stores information**

| Store # | Location | Car park[(1)] | Shop type[(2)] | Self-service %[(3)] | e-Comm operation | Number of Suppliers | Sq feet | SKUs |
|---|---|---|---|---|---|---|---|---|
| 1 | City center | None | Convenience | 68% | None | 55 | 7,192 | 5,930 |
| 2 | City center | Free on site | Convenience | 36% | Collection | 63 | 6,663 | 5,788 |
| 3 | Rural town | External paid for | Core | 49% | Collection | 155 | 23,350 | 16,810 |
| 4 | Rural town | Free on site | Core | 35% | Full e-comm | 233 | 26,540 | 17,409 |
| 5 | City center | Free on site | Core – food & home | 34% | Full e-comm | 351 | 32,032 | 18,353 |
| 6 | Rural town | Free on site | Core | 38% | Collection | 177 | 20,800 | 14,926 |
| 7 | Rural town | External paid for | Core | 32% | Full e-comm | 145 | 20,787 | 15,233 |
| 8 | Suburban | External paid for | Core – shopping center | 41% | Collection | 123 | 22,100 | 14,004 |
| 9 | Suburban | External paid for | Core | 32% | Collection | 103 | 13,560 | 9,777 |
| 10 | Suburban | External paid for | Core | 28% | Collection | 108 | 14,120 | 11,508 |
| 11 | Suburban | Free on site | Core | 43% | Collection | 144 | 19,670 | 13,986 |

*1. When a fee is required, the retailer offers a refund when payment is made via parking voucher.*

*2. "Core" are the normal supermarkets sized between 8k to 30k sq. feet, "Core – food and home" are supermarkets over 30k sq. feet, "Core – shopping center" are Core stores that operate within wider shopping centers.*

*3. Percentage of check-out lanes in the store that are self-service.*

The retailer groups the store SKUs along 27 categories of products – e.g., bakery, frozen foods, wines, etc. We extracted inventory, sales and stockcount data from the company's stock control system that included daily stock movements, product categories, and stockcount reports. For the descriptive study (presented in §3.2) and the analysis of the drivers of IRI (presented in §4), all the SKUs were included in the analysis except the ones belonging to six categories with reduced number of SKUs that grouped unique/specialized store offerings. The eliminated categories represent 1.2% of the total number of SKUs. The 21 remaining categories were aggregated into "Perishable" or "Not Perishable," eleven and ten categories respectively[2].

Across the 11 stores, the dataset included around 24,000 SKUs (the number of unique SKUs per store ranges from 5,788 to 18,353), 150,000 stockcount records and around 22 million sales records corresponding to daily sales and stock movements for each SKU in each store between the latest stockcount (available in our sample) and the immediately preceding one.

[2] The following 11 categories were considered as perishable: bakery; dairy; deli; food to go; fruit; meat; poultry, fish & eggs; prepared meals (2 categories); salads and prepared produce; vegetables. The non-perishable categories include wines; tobacco; kitchen & home essentials; health & beauty; core grocery; impulse; family & household; hospitality; beers, ciders & spirits; seasonal, occasions & gifting.

### *3.2. Measures and descriptive analysis*

First, we examine the stockcount reports of the 11 stores discussed in the previous sub-section. Such examination will inform the analysis of the IRI drivers to be discussed in §4.

The stock audit data set provides two quantities: the 'system record' defined as the inventory stock record shown in the database before the audit takes place, and 'counted record' defined as the physical stock counted during the stock audit.[3] Comparing these two quantities, we define the inventory record inaccuracy with the variable $IRI$ measuring the discrepancy between the counted record and the system record (taken as the difference of the former minus the latter). We also derive the variable $ABS_IRI$ measuring the absolute value of the IRI measure.

In addition to the data related to the physical audits conducted in all 11 stores, we consider all the inbound (replenishment) and outbound (point of sales) stock movements that took place between the latest stock audit where results are reported and IRIs are measured, and the immediately preceding one.

In more detail, this complementary dataset contains the daily replenishment quantity (if any), the end of the day stock level, the daily selling quantity, and daily sales (£), per SKU per store. The ratio between the last two measures enables us to derive the unit price per transaction for each SKU in each store. The unit price for an item is not the same in each sales transaction because of promotions. We define for each SKU the variable $PRICE$, calculated as the weighted average price over all the SKU transactions in each store. To capture promotional effects, we use a strategy to identify promotional events based on Nakamura and Steinsson (2008). We use their proposed V-shaped sales filter that records a promotion only when a price decrease is followed by a return (increase) to the prior price. Using a time window of four weeks to distinguish between regular price changes and temporary price changes, we defined the flag for a promotional discount to reflect a temporary decline in regular prices of at least 10%. The 10% threshold is commonly used in studies that distinguish promotion prices from the generally smaller changes in regular prices (e.g., Nakamura and Steinsson, 2008; Berck et al., 2008; Seaton & Waterson, 2013, Lan et al., 2022). The choice of a 4-week window is based on Nakamura and Steinsson (2008). However, we also tested the sensitivity of our findings to variations in the time window, confirming that our results remain robust within a range of 2 to 5 weeks. The variable $PROMO$ is thus defined as the total number of days an item is on temporary price promotion in each period. We focus on the duration of the discount instead of its depth to capture the effect of promotional exposure and the associated managerial oversight, conditional on other controls, which is influenced by the promotional campaign per se, and not the amount of the discount.

---

[3] Please note that the words 'count' and 'audit' are used interchangeably for the purposes of this paper. Strictly speaking, and should we wish to be consistent with the terminology used by the partnering retailer, we should be referring to 'stocktakes'. However, 'stockcounts' and 'stock audits' appear to be the more popular terms in literature and practice.

The sales data in quantity and in £ permits to derive for each SKU two variables: $QUANTITY$ and $SALES$, defined as the average daily sold quantity and sales turnover, respectively. Similarly, the data on stock positions enables us to derive the variable $STOCK$ measuring the daily average stock. We also define the variable $PER$ equal to 1 if the product belongs to a perishable category, i.e., with a relative short shelf-life measured in days, or 0 otherwise. Further, for each stock audit, we measure with the variable $DAYS$ the number of days elapsed since the immediately preceded stock audit. Finally, we define the variable $REPLEN$, measuring the number of inbound stock movements for SKU $i$ in store $s$ since the prior physical audit. To mitigate skewness and ensure comparability across variables, we log-transform all continuous variables with high variance relative to their mean – namely, *ABS_IRI, STOCK, REPLEN, QUANTITY, DAYS,* and $PRICE$ – following standard practice for count-based and scale-sensitive operational measures (Afifi & Clark, 1997; DeHoratius & Raman, 2008). Each variable was transformed as log(x + 1) prior to estimation.

Table 2 presents the descriptive statistics and pairwise correlations. IRI ranges from -873 to 1,384 units; the average magnitude of IRI for the affected SKUs is +6.6 and -5.9 units, for positive and negative discrepancies, respectively, with an average of -0.66 across all SKUs. We also find some variability of $IRI$ across stores, with the average $IRI$ ranging from -2.8 to 0.5 units.

The histogram of $IRI$ illustrated in Figure 1 reveals that 35.3% of records are accurate, 39.4% are negatively inaccurate and 25.3% are positively inaccurate. With regards to the magnitude of $IRI$, the histogram of $IRI$ shows that 90% of IRI values are between -10 and 10 units. (Although not shown in Figure 1, 11% of discrepancies have a value of -1, and 6% a value of +1.)

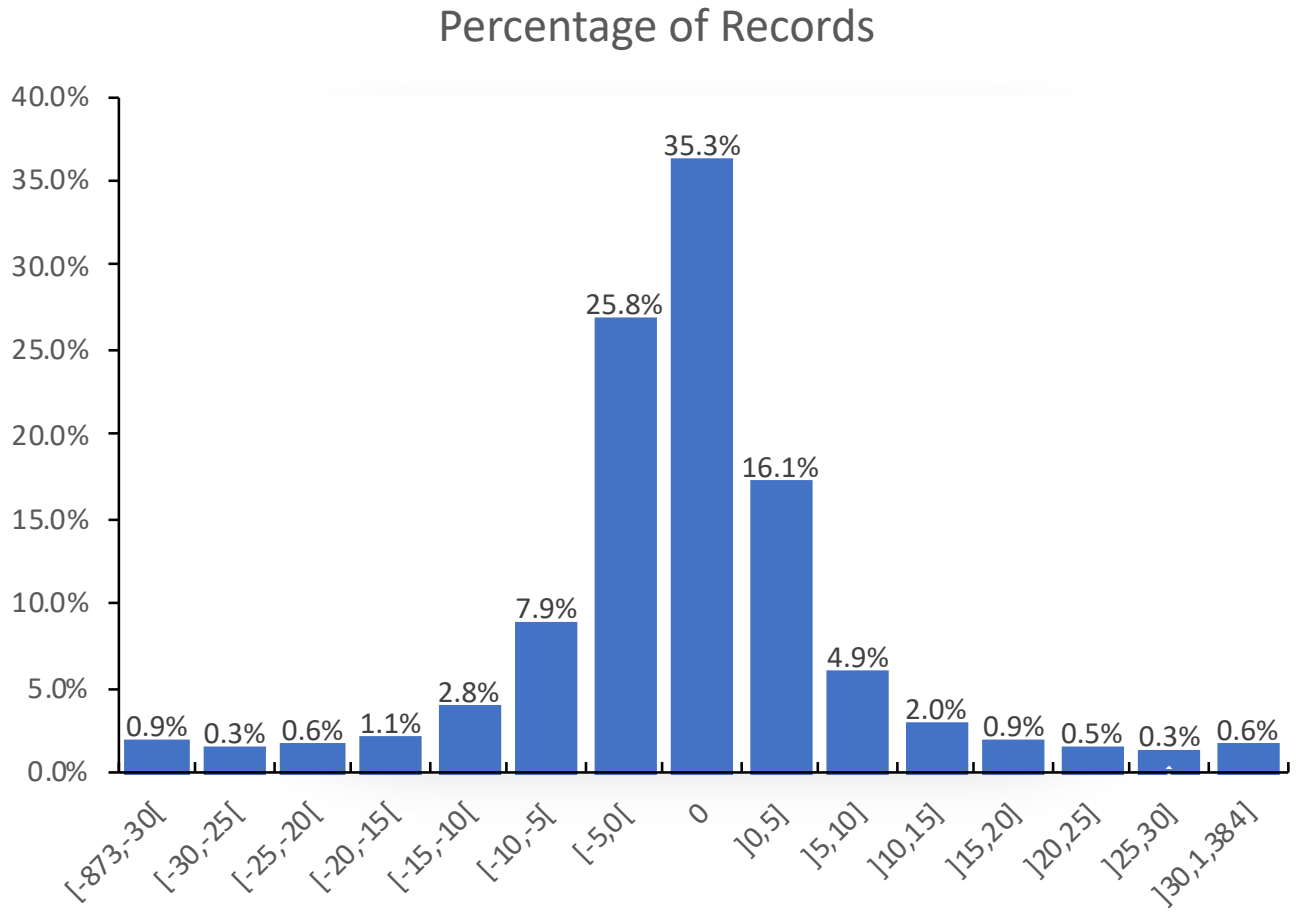

**Figure 1: IRI distribution**

The IRI distribution in our data generally matches earlier findings, both in terms of the share of SKUs affected by IRI and the direction of the discrepancy. The percentage of items with IRI (64.7%) falls within the 50 to 70% range previously reported in the literature (e.g., Kang & Gershwin, 2005; DeHoratius & Raman, 2008; Chuang et al., 2022) and the data shows a marked asymmetry with more items with negative IRI than positive (DeHoratius & Raman, 2008; Rekik et al., 2019; Ishfaq & Raja, 2020a).

**Table 2: Descriptive statistics**

**a) SKU level: n = 139,828 observations**

| Variable | Mean | Std. Dev. | Min | Max |
|---|---|---|---|---|
| *IRI* | -0.66 | 11.88 | -873 | 1,384 |
| *ABS_IRI* | 4.05 | 11.18 | 0 | 1,384 |
| *PRICE* | 3.46 | 5.13 | 0 | 160 |
| *PROMO* | 2.40 | 2.84 | 0 | 64 |
| *QUANTITY* | 1.89 | 4.97 | 0 | 288 |
| *SALES* | 3.86 | 8.49 | 0 | 377 |
| *STOCK* | 13.02 | 17.78 | 0 | 1,213 |
| *PER* | 0.25 | 0.43 | 0 | 1 |
| *REPLEN* | 20.18 | 25.59 | 0 | 248 |

**b) SKU level: pairwise correlations**

| Variables | (1) | (2) | (3) | (4) | (5) | (6) | (7) | (8) | (9) |
|---|---|---|---|---|---|---|---|---|---|
| (1) *IRI* | 1 | | | | | | | | |
| (2) *ABS_IRI* | -0.21*** | 1 | | | | | | | |
| (3) *PRICE* | 0.01*** | -0.07*** | 1 | | | | | | |
| (4) *PROMO* | -0.01*** | 0.03*** | -0.10*** | 1 | | | | | |
| (5) *QUANTITY* | 0.19*** | 0.32*** | -0.10*** | -0.05*** | 1 | | | | |
| (6) *SALES* | -0.13*** | 0.21*** | 0.09*** | -0.04*** | 0.74*** | 1 | | | |
| (7) *STOCK* | -0.26*** | 0.44*** | -0.11*** | 0.04*** | 0.53*** | 0.38*** | 1 | | |
| (8) *PER* | -0.04*** | 0.06*** | -0.06*** | -0.16*** | 0.27*** | 0.28*** | 0.01*** | 1 | |
| (9) *REPLEN* | -0.08*** | 0.13*** | -0.12*** | 0.31* | 0.38*** | 0.38*** | 0.18*** | 0.43*** | 1 |

**** p<0.001, ** p<0.01, * p<0.05*

**c) Store level: n = 11 observations**

| Variable | Mean | Std. Dev. | Min | Max |
|---|---|---|---|---|
| Average of $IRI$ | -0.78 | 1.01 | -2.76 | 0.51 |
| Average of $ABS_IRI$ | 4.10 | 1.33 | 0.97 | 5.82 |
| $DAYS$ | 151.50 | 50.90 | 91.00 | 244.00 |

### *3.3. Development of hypotheses*

We ground our hypotheses in the attention-based view of the firm (Ocasio, 1997, 2011). The ABV argues that organizational behaviour is shaped by how attention is distributed across the issues, tasks, and action alternatives competing for it. Attention is a scarce resource: decision-makers selectively notice, interpret, and act on only a subset of the operational stimuli they face. This perspective is well suited to IRI management in grocery retailing because store operations require sustained attentional engagement across a large number of SKUs, handling events, and routine execution tasks. In this setting, IRI can emerge when operational activities that require careful monitoring, verification, or system updating receive insufficient attentional engagement. Store managers and employees must divide their attention across replenishment, shelf maintenance, customer service, promotional execution, expiration date checks, and system reconciliation. Some tasks become salient and receive concentrated attention, whereas others remain routine, weakly monitored, or only indirectly visible. The ABV therefore provides a useful lens for explaining why certain operational conditions increase the likelihood that inventory records drift away from physical stock.

Three attention-based mechanisms are particularly relevant for explaining the emergence of IRI in store operations: attention scarcity, attentional selectivity, and attentional fragmentation. Importantly, these mechanisms do not operate simultaneously with equal strength; rather, different operational conditions activate distinct mechanisms, allowing us to generate differentiated predictions for each IRI driver. First, attention is scarce. Organizational actors cannot attend to all operational tasks with equal intensity at the same time, which means that some tasks receive deliberate monitoring while others are executed through routine or default behaviour (Ocasio, 1997; Rerup, 2009). Second, attention is selective. Operational tasks that are more salient, urgent, or externally visible tend to attract disproportionate attentional engagement, often at the expense of less visible but equally important activities (Simon, 1947, Kahneman, 1973; Ocasio, 2011). Third, attention can fragment across operational touchpoints. As the number of handling events, verification steps, or inventory-related interactions increases, employees must distribute their finite attention across a broader set of operational demands, reducing the depth of attention devoted to each individual task or SKU (Ocasio, 2011; Sullivan, 2010). Together, these mechanisms provide a coherent lens for predicting when inventory records are more likely to drift from physical reality. Operational conditions that increase the number of touchpoints or monitoring demands can fragment attention and increase the likelihood of IRI. By contrast, operational conditions that increase issue salience can concentrate attention and facilitate more accurate execution. In the following hypotheses, we use this framework to explain how inventory levels, replenishment frequency, perishability, and promotions shape the emergence of IRI.

Our hypotheses build on the large sample study performed by DeHoratius and Raman (2008) (DHR), who identified IRI drivers of ~10,000 different SKUs across 37 stores of a single general merchandise retailer. Building on this foundation, we examine how these drivers operate in grocery retailing, a context characterized by operational conditions that shape attentional demands in store execution. In addition to the differences in shopping behavior outlined in §2, and as discussed in the introduction, two major factors are especially salient in the grocery retailing investigated here. First, the presence of perishable products, which intensifies operational complexity through a set of mandatory shelf-management tasks (detailed in H3 below). Second, grocery sales are often enticed through promotions (Blattberg et al., 1995; Ailawadi et al., 2009). We expect those promotions to also have an impact on IRI. In the remainder of this section, we develop the hypotheses of interest to our research.

Higher average stock levels for a given SKU increase the number of operational touchpoints requiring attention: more facings to verify, more case-pack breaks to execute correctly, more backroom-to-shelf cycles to reconcile, and more opportunities for individual units to be misplaced, miscounted, or recorded inaccurately in the system (Eroglu et al., 2011; Best et al., 2022). In grocery environments, complex backroom layouts that accommodate both perishable and non-perishable items further increase the dispersion of attention across handling tasks, compounding execution errors (Gruen et al., 2002; Goyal et al., 2016). From an attention-based view, this increase in touchpoints primarily activates the *attentional fragmentation mechanism*, as employees must distribute limited attention across a larger set of interactions, reducing the likelihood of accurate execution at each step and increasing the probability that discrepancies accumulate.

Earlier research has established this relationship in general-merchandise settings (DeHoratius & Raman, 2008), but there is theoretical reason to expect it to be more pronounced in grocery retailing. In general merchandise, an additional unit of stock typically generates a limited number of handling events over its shelf life. In contrast, grocery items – particularly perishables – undergo multiple mandatory handing events between audits (see H3), each of which represents an additional attention-demanding touchpoint at which record can diverge from physical stock. As a result, the mapping between stock levels and attentional demands is amplified, strengthening the attentional fragmentation mechanism.

Taken together, these arguments indicate that higher stock levels increase the cognitive and operational load associated with inventory management by amplifying attentional fragmentation, thereby increasing the likelihood that inventory records diverge from physical stock.

*H1: IRI is positively associated with the average stock level of an item.*

While this relationship has been documented in prior research, its examination in this context allows us to assess how interactions with replenishment frequency and perishability shape the strength of the underlying attention-based mechanism. We return to these interactions below.

We next examine the role of replenishment frequency as a driver of IRI. Each replenishment cycle creates a distinct stock-handling and record-updating event: employees must verify inbound quantities, move units from backroom to shelf, place them correctly, and ensure that the corresponding system record remains aligned with the physical movement. When a SKU is replenished more frequently, it is exposed to more such episodes over the period between two audits. From an attention-based view, replenishment frequency primarily activates attentional fragmentation: accuracy depends on repeated correct execution across multiple handling and record-updating episodes, and each additional replenishment episode creates another point at which attention may be diluted or misdirected.

The operational logic reinforces this expectation. Replenishments are often carried out alongside other store tasks and under time pressure, which can amplify backroom handling complexity, shelf-placement errors, and reconciliation failures (Aastrup & Kotzab, 2009). More frequent replenishment also means more repeated interactions with both products and records, thereby increasing the number of occasions on which discrepancies can be introduced. It is therefore not surprising that replenishment activities have been identified as an important source of item out-of-stocks (e.g., Corsten & Gruen, 2003; Aastrup & Kotzab, 2009). While earlier research has suggested that replenishment frequency may trigger IRI (Ishfaq & Raja, 2020a, 2020b), the mechanism is especially salient in grocery retailing, where perishability and promotional activity generate additional mandatory handling tasks between audits. Each replenishment event in this context is embedded within a broader bundle of co-occurring operational demands, which amplifies *attentional fragmentation*: the more often a SKU is replenished, the more often the store must execute accurately a sequence of physical and informational tasks under competing attentional pressures that can generate record errors when performed imperfectly.

In view of the combined operational and theoretical arguments presented here, we expect replenishment frequency to be a significant predictor of IRI.

*H2: IRI is positively associated with the replenishment frequency.*

Note that the average inventory level and the replenishment frequency are interrelated: if the retailer decides to reduce the level of inventory kept in the store, the store needs to be replenished more frequently. The impact of the average inventory level (and the replenishment frequency) both on cost and IRI should be considered in determining order quantities for the retail store. Further, note that both product perishability and promotions can potentially shift the otherwise-optimal order quantity and frequency as the first affects how long inventory can be safely held and the second has the potential to artificially accelerate the purchase rate.

As discussed earlier, perishability introduces a set of operational challenges within the store (e.g., Ketzenberg & Ferguson, 2008; Manikas & Terry, 2010; Akkas et al., 2019). It imposes recurring mandatory attention demands – FIFO rotation, expiration date checks, markdown execution, and the removal of expired goods – that must be performed alongside routine store operations (van Donselaar et al., 2006; Akkas, 2019; Yassin & Soares, 2021). Unlike discretionary tasks, these activities cannot be postponed without immediate operational consequences and collectively generate the additional handling and record-updating events through which inventory accuracy must be maintained. From an attention-based view, perishability raises the risk of IRI because accuracy depends on sustained employee attention and repeated correct execution across more interventions, each of which creates another opportunity for products to be misplaced, quantities to be mishandled, or system records to diverge from physical stock. In particular, perishability simultaneously activates *attention scarcity* (due to competing mandatory tasks) and *attentional fragmentation* (due to repeated handling events), increasing the likelihood of execution errors. Relative to non-perishable items, perishables therefore require more mandatory interventions and generate more opportunities for execution and recording errors, increasing the likelihood of inaccurate inventory records (Chuang & Oliva, 2015).

*H3: IRI is positively associated with perishability.*

Grocery retailing is highly competitive (e.g., Colla, 2004; Cardinali & Bellini, 2014), and retailers frequently rely on promotions – either initiated by the retailer or coordinated by the manufacturer – to stimulate demand. Common instruments include price discounts, store displays, discount coupons or store flyers (see Ailawadi et al., 2009, Srinivasan et al., 2004, Tokar et al., 2011). It is not intuitively clear what the impact of these promotions on IRI will be. Promotions provide a useful setting for observing attentional selectivity because they shift managerial and employee attention toward a defined subset of SKUs for a bounded period. Under the selectivity principle, this shift reallocates attention rather than expanding the overall attention budget. It can therefore trigger two competing mechanisms with different implications for IRI, with the net effect depending on which mechanism dominates in the operational context.

The first mechanism is salience-driven *attention fragmentation*. Promotions generate demand shocks (Blattberg et al., 1995; Lam et al., 2001; Tokar et al., 2011), which increase customer interactions with promoted SKUs, accelerate stock movement, and require more frequent replenishment, display maintenance, and price changes. Although promotions concentrate attention on these SKUs, that attention may be directed primarily toward rapid promotional execution – keeping displays full, responding to demand surges, and handling customer-facing issues – rather than toward accurate record maintenance. In that case, the number of handling and recording events increases, while the likelihood of careful verification at each event declines, making IRI more likely.

*H4a: IRI is positively associated with promotional activity.*

The second mechanism emerges from *attentional selectivity*. Promotions in grocery retail are often governed by structured routines such as ticketing and price-tag compliance, display setup and reset, verification of promotional stock, and reimbursement procedures that require auditable records of promotional sales volumes (Ailawadi et al., 2009; Tokar et al., 2011). These routines are externally monitored and verification-oriented, which raises their salience relative to ordinary stock maintenance. When promotional salience is accompanied by structured checks rather than pure execution pressure, attention becomes concentrated on promoted SKUs in a way that can improve record accuracy. The same actions used to verify promotional execution may also confirm SKU identity, shelf placement, and stock-record consistency. Conditional on sales volume and stock levels, this concentration mechanism can therefore dominate the fragmentation mechanism and reduce IRI for promoted SKUs.

*H4b: IRI is negatively associated with promotional activity.*

These two hypotheses reflect a central implication of the attention-based view: while salience reallocates attention, the effect of that reallocation depends on whether it increases attentional fragmentation or induces attentional concentration through structured routines. When promotions mainly intensify execution pressure, attention fragments across a larger bundle of tasks and IRI may increase. When promotions activate structured and verification-oriented routines, attention becomes more concentrated and IRI may decrease. Support for H4b would therefore suggest that externally monitored compliance routines can convert promotional salience into record-accuracy improvements.

## 4. Drivers of IRI in grocery stores

To make our work consistent with, and be able to contribute to, the existing IRI literature – e.g., Fleisch and Tellkamp (2005), DeHoratius and Raman (2008), Hardgrave et al. (2013) – we use as a base dependent variable for this section the absolute value of IRI. While positive and negative IRI have different cost implications, we found no significant differences on their respective drivers. Accordingly, we focus our analysis on the effects of drivers on the absolute value of IRI. As a robustness test, we also replicate the analysis using a percentual measurement of IRI.

### *4.1. Controls*

In their study, DeHoratius and Raman (2008) (DHR) find that IRI is positively associated with the selling quantity of an item (h1) and negatively associated with the frequency of audits (h2), the cost of an item (h3), and the dollar volume of an item (h4).[4] DHR also find drivers at the store level, namely, that IRI is associated with the distribution structure used (h5), and that it is positively associated with the inventory density (h6) and the product variety (h7) within the store. We exclude from our study the storewide elements identified by DHR (h5-h7) as our sample does not permit sufficient variance across stores. We control for all time-invariant store characteristics through a store-fixed effect in our statistical models. As such, our research focuses on SKU-level attributes as drivers of IRI as opposed to the mixed approach followed by DHR that had enough store variance to capture store-level attributes. Finally, it should be noted that our research partner does not perform partial audits of targeted SKUs, relying instead on storewide stockcounts. Thus, our measures of frequency of testing and the store-fixed effects are collinear. We explicitly control for this in the estimation section.

### *4.2. Empirical estimation*

We used a fixed effects (FE) model to control for time-invariant store characteristics and account for potential differences among SKUs by estimating the model with random effects (RE). Specifically,

$$\begin{aligned}|IRI_{is}| = \beta_0 + \alpha_1 STOCK + \alpha_2 REPLEN + \alpha_3 PER + \alpha_4 PROMO + \gamma_1 QUANTITY + \gamma_2 DAYS \\ + \gamma_3 PRICE + \boldsymbol{\eta_s} + \boldsymbol{\kappa}_{\mathbf{i}} + \varepsilon_{is}\end{aligned}$$

where $IRI_{is}$ is the inventory record inaccuracy of item *i* in store *s*. $\beta_0$ is a fixed intercept parameter, $\boldsymbol{\eta_s}$ is a vector with the store's fixed effects, and $\boldsymbol{\kappa_i}$ is a vector with the SKU random effects. The *QUANTITY, DAYS* and *PRICE* regressors are controls introduced as part of the drivers already identified by DHR (h1 through h3). A control for the sales volume (in dollars/pounds) of an item (h4) was dropped from the regression since margins for all products and the sales volume are highly collinear with the *QUANTITY* and *PRICE* controls. The $\alpha$ coefficients allow us to test our four hypotheses.

---

[4] We use lower case h to designate the hypotheses from the DHR study. Hypotheses from our study are designated with capital H.

Model 1 in Table 3 reports the results of a regression with an intercept, store-fixed effects, and SKU random effects, with robust standard error in parentheses. The variance across stores explains 10% of the observed variance in the IRI. Model 2 reports the results of our regression. The SKU-level variables, including controls, explain an additional 17% of the observed variance in IRI. Note that the three controls introduced as part of DHR's previous findings are all significant (*p*=0.000 for $\gamma_1, \gamma_2,$ and $\gamma_3$) and have the expected sign. This confirms that despite the differences in the retail environment, DHR's findings hold, and volume, frequency of audits, and price are strongly associated with IRI.

Note, however, that in the Model 2 regression, the intercept has been dropped as, as explained above, all the items in a store share the same number of *DAYS* since the last audit, i.e., is colinear with the store-fixed effects. Indeed, Model 2 has a mean Variance Inflation Factor (VIF) of 9.2 with 5 regressors (*DAYS* and 4 store FE) having a VIF higher than 10. We confirm the source of collinearity by dropping the control for the *DAYS* since last inspection (Model 3). The revised model has a maximum VIF of 4.25 (mean VIF = 2.05). The revised specification reintroduces the intercept and all coefficients and standard errors remain unchanged, indicating that all information in the *DAYS* variable reverts to the store-fixed effects. The fact that *DAYS* has a significant coefficient in the presence of store FE indicates that it contains useful directional information for the explanation of IRI. Note, however, that the inclusion of *DAYS* was to control for one of the pre-existing determinants found by DHR (h2) and it is not central to our hypotheses testing. We, therefore, report Model 2 to retain the directional information contained in *DAYS* but treat Model 3 as our preferred, more parsimonious specification.

Regarding re-stocking practices, consistent with H1, we find evidence that higher average inventory levels are indeed associated with higher IRI ($\alpha_1 = 0.521,$ *p*=0.000). The frequency of store replenishments (H2) is also highly significant and has the predicted sign ($\alpha_2 = 0.056,$ *p*=0.000). H3 predicts that perishable products will exhibit higher IRI within the store. We find, in support of this hypothesis, that the coefficient of the *PER* indicator is highly significant ($\alpha_3 = 0.064,$ *p*=0.000).

Finally, regarding the effect of promotions on IRI, we find that the fraction of items sold under promotion is negatively associated with IRI ($\alpha_4 = -0.012,$ *p*=0.000), thereby rejecting H4a and in support of H4b. That is, the extra operational oversight triggered by preparing for and monitoring promotions is associated with lower IRI when controlling for the negative effects of incremental sales on IRI.

**Table 3: IRI drivers**

| MODEL<br>Dependent var. | (1)<br>ABS_IRI | (2)<br>ABS_IRI | (3)<br>ABS_IRI | (4)<br>ABS_IRI% |
|---|---|---|---|---|
| $\beta_0$ CONSTANT | 0.878*** | | -0.333*** | 0.010 |
| | (0.007) | | (0.016) | (0.007) |
| $\alpha_1$ STOCK (+) | | 0.521*** | 0.521*** | 0.083*** |
| | | (0.006) | (0.006) | (0.002) |
| $\alpha_2$ REPLEN (+) | | 0.056*** | 0.056*** | 0.011*** |
| | | (0.004) | (0.004) | (0.002) |
| $\alpha_3$ PER (+) | | 0.064*** | 0.064*** | 0.024*** |
| | | (0.008) | (0.008) | (0.003) |
| $\alpha_4$ PROMO (±) | | -0.012*** | -0.012*** | -0.002 |
| | | (0.003) | (0.003) | (0.001) |
| $\gamma_1$ QUANTITY (+) | | 0.038*** | 0.038*** | 0.008* |
| | | (0.009) | (0.009) | (0.004) |
| $\gamma_2$ DAYS (–) | | -0.066*** | | |
| | | (0.003) | | |
| $\gamma_3$ PRICE (–) | | -0.073*** | -0.073*** | -0.008*** |
| | | (0.005) | (0.005) | (0.002) |
| Store FE | Yes | Yes | Yes | Yes |
| SKU RE | Yes | Yes | Yes | Yes |
| | | | | |
| Observations | 139,830 | 139,830 | 139,830 | 137,442 |
| Pseudo $R^2$ | 0.097 | 0.268 | 0.268 | 0.082 |

*The Pseudo $R^2$ measure is based on the squared correlation between the estimated and actual orders (Wooldridge 2010). Robust standard errors are in parentheses*
**** p<0.001, ** p<0.01, * p<0.05*

It is interesting to note that the effect of the *STOCK* level on IRI is an order of magnitude greater than the effect of all the sales *QUANTITY* – a 1% increase in the stock level represents a 0.52% increase in IRI, while a 1% increase in the sales quantity only accounts for a 0.04% increase in IRI – and that the effect of replenishment frequency (*REPLEN*) is 47% higher than the effect of *QUANTITY.* These two results alone provide a more nuanced insight into the drivers of IRI than the study by DHR that included only the sales quantity. Regarding our product attributes, perishable items within the store (*PER*) have, on average, a 6.4% greater record inaccuracy than non-perishable items, and items under promotion (*PROMO*) have a 1.2% lower IRI than not-promoted items.

Models 1 through 3 have as a dependent variable the absolute value of IRI. In Model 4, as a robustness test, we use as a dependent variable the absolute value of IRI expressed as a percentage of the inventory level at the time of the audit. Our sample size drops by 1.7% as some items with IRI had no actual inventory at the time of the audit, and, not surprisingly, the variance explained by the model drops as our dependent variables were designed to explain the absolute value of IRI and not its percentage value. Nevertheless, the significance and sign of most coefficients of interest is the same as in Model 3. The only discernable difference is the drop in significance of the sales *PROMO* ($p$ = 0.105). This drop in significance is explained by the fact that the percentage dependent variable adjusts for the relative magnitudes causing more than a 50% drop in the $z$ values of most regressors.

## 5. Do audits improve sales?

We now turn our attention to the extent to which IRI monetarily matters in retailing. Do audits and inventory record reconciliations improve sales? To evaluate this question, we performed a quasi-experimental intervention study (Shadish et al., 2002), comparing the performance of two similar stores where the 'treatment' consisted of a store-wide stocktake in one of the stores, approximately six months after an initial stocktake in both stores. We were particularly interested in assessing whether the re-ordering and cleaning of the store had an impact on aggregate store sales. As discussed above, this experimental design also allowed us to assess heterogeneous effects of audits on products depending on their attributes or their accuracy state. The retailer assisted in the matching of treatment and control stores based on similarities in size, location, format, product assortment, pricing, staffing levels, and replenishment and re-shelfing procedures. Although the control store had more advanced e-commerce functions, both stores exhibited similar pre-treatment sales trends and organizational structures. A *parallel trends* test was conducted to validate comparability prior to the audit intervention. Once the matched stores were identified, we confirmed with the retailer that policies and organizational capabilities in these stores were not changed for the duration of the experiment. The selection of the treatment store was done randomly through the flip of a coin. Specifically, store #1 in Table 1 served as our treatment store, and store #2 served as a control. The design maximizes internal validity by analyzing more than 5,000 SKUs observed over multiple months, providing substantial within-store variation and enough statistical power. To our knowledge, no prior empirical study has implemented a store-wide field experiment to assess the direct implications of inventory record inaccuracy for sales performance, making this investigation a rare and valuable test of theory in a natural retail setting.

Sales and inventory data were tracked for two[5] months before and after the treatment. For our analysis, we only used SKUs sold before and after the stock count treatment in both the treatment and control stores. This resulted in a sample of 5,657 directly comparable SKUs, representing 95% and 97% of the total number of SKUs in the treatment and control stores respectively. We use the monthly data aggregation following previous work in retail operations (e.g., Ton & Raman, 2010; Kesavan et al., 2014; Chuang & Oliva, 2015). We present our analysis in sales revenue, but all results are consistent when performing the analysis measuring sales in units.

Note that controlling the sample to include only matching SKUs allows us to omit category and SKU effects from our regressions – being time- and store-invariant, their effects are captured in the higher-level fixed effects. Although incorporating nested random effects for category and SKU (three level model) does reveal some residual interclass correlation, their inclusion does not affect the estimates from the models reported here other than yielding slightly tighter standard errors. Since our treatment is established at the store level, we proceed with presentation and interpretation of the more parsimonious models.

---

[5] The two-month pre- and post-treatment windows were organizational constraints and defined to minimize the interfering with normal operations for the retailer, e.g., withholding treatment to the control store for too long.

This experimental design allowed us to perform rigorous diagnosis of the experimental conditions and assumptions, and detailed analysis of the effect of the intervention across all the store SKUs and specific heterogenous effects on product attributes and inventory state. To reinforce the credibility of the research design, we complement the pre-treatment assessment with formal diagnostics: (i) statistical tests confirming pre-treatment comparability, (ii) a placebo month (–1) showing no anticipatory effect, and (iii) a dynamic Difference-in-Differences (DID) specification indicating post-treatment gains that strengthen in month +2. These checks follow recommended best practices for validating the parallel-trends assumption in DID analyses (Angrist & Pischke, 2009; Bertrand et al., 2004; Cunningham, 2021) – see also Figure 2 for visual confirmation of the dependent variable trajectories by store.

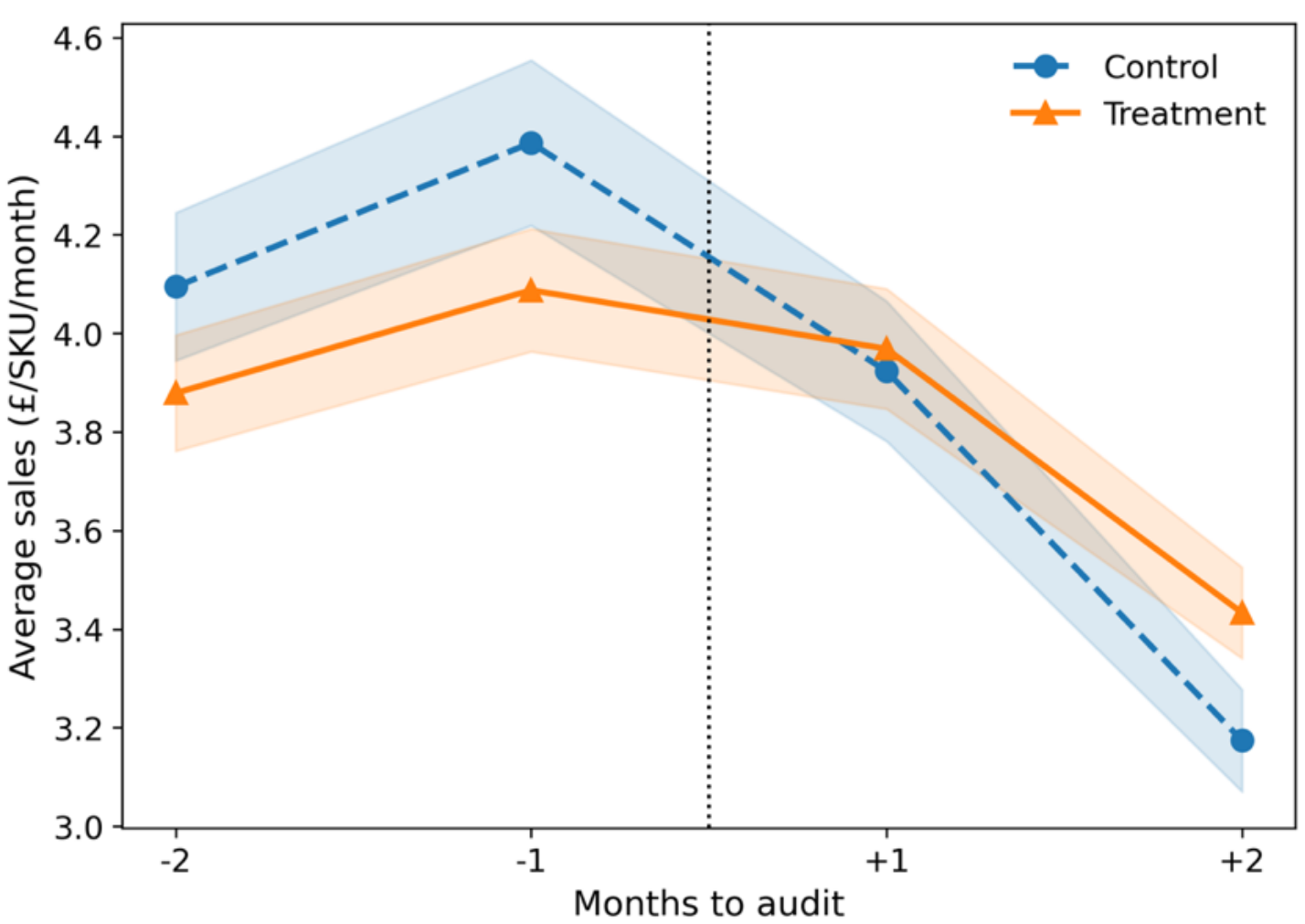

**Figure 2: Monthly average sales per SKU before and after treatment**
(shaded areas represent the 95% CI of the reported mean)

In what follows, we first assess the mean treatment effect using a DID model (Model 1); we separate the monthly fixed effects (Model 2) to test for the parallel trends and the Granger causality test; and in Model 3 we formulate a dynamic specification of the treatment effect to assess effect of the treatment over time. Finally, we assess the heterogenous effects of the treatment on whether an item was corrected and the sign of the correction (Model 4) and on product perishability (Model 5); a key driver of IRI in the grocery environment. Throughout this progression of sophistication of tests and analysis, results are consistent within and across models, thus further establishing the robustness of our results.

### *5.1. Mean treatment effect*

We first assess the standard DID model motivated by Card and Krueger (1994) that addresses the possibility of unobserved variation between the treatment and control groups, as well as exogenous temporal effects. The specification of our base regression model is:

$$y_{ist} = \beta_0 + \beta_1 POST_t + \beta_2 TREAT_s + \gamma POST_t \times TREAT_s + \varepsilon_{ist} \quad (1)$$

where $y_{ist}$ is the sales of SKU *i* sold in store *s* in month *t*. $POST_t$ is a period binary indicator (1 if month *t* is in the *post* treatment period) and the estimate $\beta_1$ captures the sales variation across the two periods (*pre* vs. *post*). $TREAT_s$ is a store indicator (1 if store *s* is the *treatment* store) and the estimate of $\beta_2$ indicates the intrinsic difference in sales metrics between the treatment and the control store. The parameter $\gamma$ is the DID estimator of interest that captures the underlying impact of the audit after isolating for store and period-fixed effects. We adopt robust standard errors (Cameron & Miller, 2015) clustered by SKU to control for SKU-specific effects.

Model 1 in Table 4 shows the regression results. The significance of the $\beta_1$ coefficient indicates a general reduction of sales (across both stores) during the post-audit period, and the significance of the $\beta_2$ estimate suggests a slight difference between the treatment and control store (the treatment store had on average 6% lower sales). The positive and significant estimate of the DID coefficient ($\gamma = 0.396$, *p*=0.000) implies that the audit had a positive effect on sales in the treatment group. The DID estimator targets the treatment contrast rather than maximizing explained variance; therefore, low $R^2$ values (near zero in Table 4) are not, by themselves, diagnostics of bias (Wooldridge, 2010; Angrist & Pischke, 2009; Cunningham, 2021). That said, the low fit reflects the high idiosyncrasy of SKU-month sales and the fact that our hypothesized factors are only a small fraction of all the drivers of IRI (i.e., staffing, theft, store operations, etc.), and underscores our cautious interpretation of the results, as well as the value of using robust standard errors and multiple specification checks. The DID sales uplift estimate (£0.40) represents 11.2% of the post-audit average sale per item-store (£3.53).

An alternative model (Card, 1992) that isolates the time-invariant store-specific characteristics (e.g., square footage and density of competitors in the neighborhood) and each month-fixed effect (e.g., holidays and weather patterns) is specified as follows:

$$y_{ist} = \beta_0 + \eta_s + \boldsymbol{\mu_t} + \gamma POST_t \times TREAT_s + \varepsilon_{ist} \quad (2)$$

**Table 4: Difference-in-Differences (DID) analysis**
**(treatment store: #1, control store: #2; see Table 1)**

| MODEL<br>Dependent Variable | (1)<br>avg_sales | (2)<br>avg_sales | (3)<br>avg_sales | (4)<br>avg_sales | (5)<br>avg_sales |
|---|---|---|---|---|---|
| $\beta_0$ CONSTANT | 4.210*** | 4.079*** | 4.359*** | 4.079*** | 2.181*** |
| | (0.151) | (0.141) | (0.166) | (0.141) | (0.103) |
| $\beta_1$ PERIOD (POST=1) | -0.685*** | | | | -0.297*** |
| | (0.044) | | | | (0.034) |
| $\beta_2$ TREAT | -0.250** | | | | -0.043 |
| | (0.078) | | | | (0.071) |
| $\beta_3$ PER | | | | | 6.407*** |
| | | | | | (0.414) |
| $\gamma$ DID – PERIOD x TREAT | 0.396*** | 0.396*** | | -0.059 | 0.204*** |
| | (0.040) | (0.040) | | (0.145) | (0.036) |
| $\gamma_{-2}$ DID x MONTH$_{-2}$ | | | 0.082 | | |
| | | | (0.052) | | |
| $\gamma_1$ DID x MONTH$_1$ | | | 0.329*** | | |
| | | | (0.052) | | |
| $\gamma_2$ DID x MONTH$_2$ | | | 0.545*** | | |
| | | | (0.060) | | |
| $\delta_p$ POS | | | | -0.129 | |
| | | | | (0.262) | |
| $\delta_n$ NEG | | | | 1.111*** | |
| | | | | (0.225) | |
| $\eta_s$ STORE FE | | -0.250** | -0.291** | -0.250** | |
| | | (0.078) | (0.089) | (0.078) | |
| $\mu_{-2}$ MONTH FE | | | -0.301*** | | |
| | | | (0.060) | | |
| $\mu_{-1}$ MONTH FE | | 0.259*** | | 0.259*** | |
| | | (0.047) | | (0.047) | |
| $\mu_1$ MONTH FE | | -0.242*** | -0.468*** | -0.242*** | |
| | | (0.045) | (0.047) | (0.045) | |
| $\mu_2$ MONTH FE | | -0.869*** | -1.203*** | -0.870*** | |
| | | (0.054) | (0.080) | (0.054) | |
| $\alpha_1$ PER x PERIOD | | | | | -1.258*** |
| | | | | | (0.123) |
| $\alpha_2$ PER x TREAT | | | | | -0.652** |
| | | | | | (0.204) |
| $\theta$ DID x PER | | | | | 0.606*** |
| | | | | | (0.106) |
| Observations | 45,110 | 45,110 | 45,110 | 45,110 | 45,110 |
| Adj $R^2$ | 0.001 | 0.001 | 0.001 | 0.002 | 0.079 |
| F test | 89.47 | 65.14 | 58.74 | 57.04 | 73.72 |

*Robust standard errors are in parentheses*
**** p<0.001, ** p<0.01, * p<0.05*

where $\eta_s$ is a dummy for one of the stores (in this case the treatment store), and the vector $\boldsymbol{\mu_t}$ are the coefficients for the fixed effects for each month in the sample. This formulation slightly improves the model fit but yields an identical estimate for the DID estimator $\gamma$ (Model 2 in Table 4), thus providing corroboration of our main finding. Augmenting Model 2 to include the 3-way interaction between *MONTH×TREAT* and variables indicating pretreatment and posttreatment periods, reveals that there is no difference in slopes between the treatment group and the control group ($p$=0.112), i.e., the *parallel trends* test (Cunningham, 2021)[6], while indicating a significant difference in slopes in the posttreatment periods. Furthermore, Model 2 reveals that most of the sales drops (across both stores) occurred during the second month after the intervention ($\mu_2 = -0.869$, *p*=0.000). Note that the coefficients of the month base case represent adjustments to the sales levels two months prior to the inventory audit (the base case).

### *5.2. Dynamic effects*

Next, we test for the dynamics of the treatment effect as the benefits from an inventory audit, and the re-arrangement of shelves that it implies, are expected to decay over time. We formulate the dynamic specifications for DID as follows:

$$y_{ist} = \beta_0 + \eta_s + \boldsymbol{\mu_t} + \boldsymbol{\gamma_t} \times \boldsymbol{TREAT_{st}} + \varepsilon_{ist} \quad (3)$$

where the vector $\boldsymbol{TREAT_{st}}$ is a binary indicator of whether store *s* had been treated in month *t.* For the estimation, we omit the pre-intervention period (month= –1) to leave it as a reference category. As expected, the coefficient for the remaining month prior to the intervention ($\gamma_{-2}$) is not significant (see Model 3 in Table 4), confirming that the treatment effects cannot be observed in anticipation of the treatment – this is the Granger-type causality test (Angrist & Pischke, 2009).[7]

The coefficients of the two periods after the intervention do show the expected improvement in sales, but contrary to expectations, we observed that the sales uplift during the second month after the intervention is 65% higher than during the first month. This could be explained in part by the disruptive effect that the audits might have on store operations and traffic (thus reducing sales during the first few days after the audit) and by the cumulative effect of audits on traffic (an ordered store with less stockouts will attract more traffic), or by the fact that sales for the second period after the audit are much lower for both stores (see the coefficient for $\mu_2$ in Models 2 and 3), thus making the benefits of the audit more salient.

### *5.3. Heterogenous effects – IRI correction*

We test for the possibility of heterogeneous effect of the direction of the inventory correction performed during the treatment audit by creating binary variables capturing the sign of the correction performed (*POS=1* if the actual inventory was found to be greater than the system inventory and *NEG=1* if the actual inventory was less than the system record inventory), with no sign assigned if the item was audited and detected as accurate. The base case, an accurate item, allows us to isolate the effect of the *intent to*

---

[6] We confirm this result using the *didregress* command in Stata and obtaining the formal test of parallel trends (*estat pretends*). Model 2 also passes the Granger causality test (*estat granger*) (*p* = 0.112), described in detail for Model 3.

[7] Model 3 also passes the parallel trends test as implemented in Stata (*p* =0.114).

*treat* but when no actual treatment was necessary (Gupta, 2011; Zhang et al., 2017), thus allowing to test whether there are store-wide effects of the audit beyond the corrected items. We specified this model based on Model 2 that explicitly controls for the store- and month-fixed effects:

$$y_{ist} = \beta_0 + \eta_s + \boldsymbol{\mu_t} + \delta_p POS + \delta_n NEG + \gamma POST_t \times TREAT_s + \varepsilon_{ist} \quad (4)$$

Note that in this specification the *POS* and *NEG* dummies are only defined for the *POST*×*TREAT* condition and thus embed the three-way interaction. By the same logic, the two-way interactions {*POS*×*POST, POS*×*TREAT, NEG*×*POST, NEG*×*TREAT*} are collinear with these terms and dropped from the regression. Model 4 in Table 4 reports the regression results. First, note that the coefficients of the monthly fixed effects are the same as in Model 2, thus all differences are centered in the difference-in-differences term and the direction of correction dummies. When separating the effect of the correction sign, the base case, that is an item audited but not corrected, the difference-in-differences coefficient ($\gamma$) is no longer significant ($p$=0.786), indicating that there are no extended benefits, as measured by increases in sales, of performing an audit beyond the corrected items. The coefficient of the negative corrections ($\gamma + \delta_n = 1.052$) is positive and highly significant ($p$=0.000) whereas the coefficient of the positive corrections ($\gamma + \delta_p = -0.188$) is not significantly different from zero ($p$=0.427), suggesting that the incremental sales benefits (the dependent variable of interest) of the audit process are only realized in the items that are overreported in the information system. Indeed, the treatment effects in items that required a negative correction is 2.67 times the average effect across all items in the treatment store and 32% of the post-audit average sale per item-store[8]. This is consistent with the "inventory freezing" scenario (Kang & Gershwin, 2005) where, as previously discussed, an OOS item is not replaced because the system record is positive, despite no stock being available. The absence of effect on sales of the items that were audited but not corrected ($\gamma$) or that were positively corrected ($\gamma + \delta_p$) suggests that the simple process of re-sorting the store is not enough to have a positive effect on sales. Thus, although the storewide audit could have some marginal operational benefits, our analysis shows that these alone do not translate into measurable sales increases. The absence of significant effects for positively corrected or uncorrected items confirms that it is the correction of overstated inventory records that drives the observed sales uplift.

---

[8] The factors are calculated comparing the treatment effect on items that required a negative correction in Model 4 ($\gamma + \delta_n$) to the average treatment effect in Model 2 ($\gamma$) and the post-treatment average sales in Model 1 ($\beta_0 + \beta_1 + \beta_2$).

### *5.4. Heterogenous effects – perishability*

Finally, to test for the possibility of a heterogeneous treatment effect on one of the key dimensions affecting IRI (see §3), we separate the effects of the audits on perishable products vs. non-perishable products. We include in the perishable product category those SKUs that have a short (few days) expected shelf life, leaving in the non-perishable category products with expected shelf life measured in months (see footnote 2 for detailed list of categories classified as perishable). For our sample stores, perishable SKUs represent 27.5% of the total number of SKUs in the store. We expand the model in equation 1 to include the triple-difference model that includes all two-level interactions:

$$y_{ist} = \beta_0 + \beta_1 POST_t + \beta_2 TREAT_s + \beta_3 PER_i + \alpha_1 PER_i \times POST_t + \alpha_2 PER_i \times TREAT_s + \gamma POST_t \times TREAT_s + \theta PER_i \times POST_t \times TREAT_s + \varepsilon_{ist} \quad (5)$$

In the above model, $PER_i$ is an indicator variable that measures whether SKU $i$ is perishable and whether the estimate of $\theta$ captures the incremental effect the audits will have on perishable products over the effect calculated for non-perishable products ($\gamma$). Model 5 in Table 4 reports our regression results. First, it is interesting to note that when controlling for perishable products, the difference between the treatment store and the control store becomes non-significant ($p$=0.765 for $\beta_2$), suggesting a significant difference in the volume of perishables sales across the two stores. The coefficient for perishable ($\beta_3$) is positive and significant indicating overall higher average sales of products in the perishable categories. Interestingly, the effect of the audits on sales of perishable products ($\theta$=0.606) is almost three times higher than for the non-perishable products ($\gamma$=0.204), suggesting that audits can play an important role in improving sales of those products with the highest inventory record inaccuracies that we identified in §3.

## 6. Discussion

Understanding how operational processes affect performance remains critical for retail firms operating in mature and highly competitive markets. In grocery retailing, inventory management is especially complex because perishability and promotional activity shape the frequency, visibility, and execution of inventory-related tasks. Against this background, this paper examined the prevalence and drivers of IRI and assessed the impact of store-wide audits on sales performance using data from a leading UK grocery retailer. The discussion below distinguishes theoretical, empirical, and managerial implications of these findings before outlining the main limitations of the study and directions for future research.

### *6.1. Theoretical implications*

Our findings extend the attention-based view (ABV) in operational settings (Ocasio, 1997, 2011; Joseph et al., 2024) by showing how different inventory conditions activate distinct attention-allocation mechanisms that shape the emergence, visibility, and correction of inventory record inaccuracies. More broadly, our results refine the ABV by showing that operational accuracy outcomes are not driven solely by attention scarcity, but by the interplay between attentional fragmentation and attentional selectivity (Ocasio 1997, 2011). While prior work emphasizes limited attention as a constraint, our findings show that certain operational conditions – such as promotions – can temporarily improve accuracy by concentrating attention through structured routines (Rerup, 2009).

First, building upon foundational work by DeHoratius and Raman (2008), we empirically validate additional operational drivers of IRI that are especially salient in grocery retailing – namely, product perishability and promotional activities. Higher stock levels, more frequent replenishment, and perishability are all associated with higher IRI because they increase the number of handling, verification, and record-updating episodes through which accuracy must be maintained. From the perspective of the ABV, these conditions multiply the touchpoints at which errors can be introduced and therefore increase the scope for attention to fragment across repeated interventions (Ocasio 2011; Sullivan, 2010). Our finding that promotional activity is associated with lower IRI is particularly informative in this regard. Our results suggest that, in this context, structured promotional routines drive attentional selectivity with attention concentrated on promoted items. This attention concentration dominates over fragmentation effects**,** leading to improved record accuracy. The contribution here is therefore not simply to identify additional IRI drivers, but to refine the attention-based view by showing that operational conditions differ systematically in how they shape attention: some fragment attention across repeated touchpoints, whereas others concentrate attention through salience and structured verification (Joseph et al., 2024).

Second, our audit findings further refine the ABV by showing that the performance consequences of IRI depend not only on whether inaccuracies exist, but also on whether they are visible to routine organizational processes. The sales benefits of audits are concentrated in items with negative IRI, which suggests that audits matter most when they surface discrepancies that would otherwise remain hidden from normal replenishment routines. In that sense, the audit acts not simply as a counting device, but as a visibility-restoring intervention that makes overstated records observable and therefore correctable (Ocasio 1997; Vuori & Huy, 2016). The stronger audit response among perishable items is consistent with the same logic: where handling intensity is greater and inaccuracies are more prevalent, the returns to restored visibility are correspondingly larger. This extends the ABV by showing that attention matters not only in the generation of operational errors, but also in the conditions under which those errors become sufficiently visible to trigger corrective action (Ocasio 2011; Joseph et al. 2024).

Third, the temporal pattern of the treatment effect suggests that the benefits of attention-directing interventions may unfold gradually rather than dissipate immediately. This pattern is consistent with the possibility that audits do more than correct records in a single moment; they may also temporarily recalibrate operational routines, allowing the downstream sales consequences of improved availability to emerge over time. From an ABV perspective, this opens an important avenue for future research on the persistence of attention, the durability of corrective routines, and the timing of interventions in operational settings (Ocasio, 2011; Rerup, 2009).

Finally, we note that our proposed hypotheses have theoretical generalizability (Meredith, 1998). While the empirical findings from a field quasi-experiment lack sample generalizability – see discussion in §6.4 – the identified drivers and dynamics are likely to extend to other retail formats facing similar inventory challenges and operational tradeoffs, including convenience stores, drugstores, and apparel retailers. The specific magnitude of the effects and trade-offs will differ from setting to setting, but the framing and methods proposed here open a new line of research. In this case, it is the ecological realism of the field-experiment, that captures managerial behavior, operational routines, customer flows, and system frictions in the environment where the phenomenon naturally occurs, that gives credence to the identified causal relationships (McGrath, 1981).

### *6.2. Empirical contribution*

This study also makes three distinct empirical contributions to the literature on inventory record inaccuracy and retail operations. First, we implement a store-wide quasi-experimental design that provides direct causal evidence on the impact of inventory audits on sales performance at scale. Rather than focusing on a handful of SKUs or relying solely on simulation studies, we analyze more than 5,000 SKUs across a matched pair of stores using a difference-in-differences framework with store and month fixed effects, formal parallel-trends diagnostics, and a dynamic specification that traces treatment effects over time. This design offers a rare field-based view of how audit interventions translate into sales outcomes in a natural retail setting.

Second, our results yield empirical regularities that can serve as quantitative benchmarks for future work. We estimate that a store-wide stockcount is associated with approximately an 11% increase in sales over the two months following the audit, providing an order-of-magnitude assessment of the economic value of correcting inventory records. We further show that this uplift is highly heterogeneous. The benefits are concentrated on SKUs with negative IRI at the time of the audit, while SKUs with positive IRI or accurate records exhibit no discernible sales response. Perishable items – which also exhibit higher baseline IRI – experience a stronger post-audit sales improvement than non-perishable items. Together these findings sharpen our empirical understanding of where and for whom audits matter most and provide quantitative reference points for subsequent studies.

Third, we document dynamic treatment effects that enrich the empirical foundation for future work on audit timing and targeting. The treatment effect strengthens in the second month after the audit rather than fading immediately, suggesting that the benefits of record correction may unfold gradually as replenishment routines and customer demand adjust. The decomposition of effects by IRI sign and product perishability also highlights specific product groups – especially perishable SKUs with overstated records – as promising targets for more selective audit policies. These empirical patterns can help guide the calibration of future models and the design of subsequent intervention studies.

These empirical patterns also have direct implications for how retailers should allocate audit effort, managerial attention, and supporting technologies.

### *6.3. Managerial implications*

Our findings provide actionable guidance for retail managers seeking to improve inventory accuracy and sales performance. First, the identification of stock levels, replenishment frequency, and perishability as key drivers of IRI can inform the prioritization of inventory-control effort. Products exposed to frequent handling, repeated replenishment, and stricter shelf-life management are more likely to experience faster information-degradation rates (Chuang et al., 2022) and should therefore receive greater monitoring attention, more disciplined execution routines, and, where appropriate, selective technological support.

Second, the heterogeneous audit effects imply that retailers need not treat all SKUs as equally valuable targets for inventory correction. Because the sales benefits of audits are concentrated among items with negative IRI and are stronger for perishable products, retailers can increase the return on inventory-control investments by prioritizing categories and items with greater risk of hidden stock discrepancies. This points toward a more selective audit logic in which frequency, labor, and follow-up attention are allocated according to product characteristics and discrepancy risk rather than uniformly across the store. While we did not have access to the cost of performing an audit, the method described here allows managers to quantify the benefits of an audit and make a more informed decision on the timing and intensity of future audits. The empirical demonstration of an approximate 11% store-wide sales increase following inventory audits suggests that audits should be viewed as valuable operational investments, shifting their perception from compliance-focused tasks to a strategic business driver.

Third, the negative association between promotions and IRI shifts the calculus of the trade-offs between the benefits of promotions and the disruptions they cause to regular operations. In the focal retailer, promotional campaigns appear to introduce structured routines and heightened scrutiny that reduce discrepancies, suggesting that similar practices could be extended beyond promotional periods. Activities such as tighter shelf checks, more frequent stock verification, clearer task ownership, and stronger execution discipline during promotions may serve as templates for accuracy-enhancing routines in day-to-day operations.

Finally, the findings have implications for the role of technology in inventory management. Prior work shows that technologies such as RFID can improve inventory visibility and reduce IRI, but they do not eliminate the problem entirely (Hardgrave et al., 2009, 2013; Goyal et al., 2016). Our results reinforce this view by showing that many discrepancies arise from execution and handling processes that still require human interpretation and corrective action. Technologies that improve visibility should therefore be seen as complements to, rather than substitutes for, managerial oversight and targeted audits. Hybrid approaches that combine selective technology deployment with focused audits and disciplined operating routines are likely to be most effective in improving inventory record accuracy and associated sales performance.

### *6.4. Limitations and future research*

Like any empirical study, this work has limitations. Field experiments, in general, achieve greater realism by sacrificing contextual generalizability and full control of the experimental setting (McGrath, 1981, Abdulla et al., 2024). In our case, the quasi-experimental study was conducted on one matched pair of stores within a single retailer, and economic constraints limited the degree to which the control conditions could be held completely stable. Further tests are need to verify whether stockcounts can have a generally positive effect on sales by decluttering the store and the directionality and scale of the heterogenous effects detected here, including those associated with IRI status, perishability, and promotions. Replication across diverse retail settings and larger samples would strengthen external validity. The quasi-experimental structure proposed here can be adapted by retailers seeking to assess the sales impact of inventory corrections in their own operations. For example, the mechanisms underlying our finding that promotional activity is associated with lower IRI may depend on retailer-specific promotional routines, monitoring practices, and compliance structures. In the focal setting, promotions appear to trigger strengthened monitoring and compliance processes that reduce discrepancies, but retailers that allocate less operational attention to promotional products may not experience the same effect. Future research should examine how variation in promotional governance shapes the relationship between promotions and inventory record accuracy.

The study is also subject to limitations related to unobserved managerial heterogeneity. Although store fixed effects control for time-invariant differences across stores, and we attempted to control for changes in managerial processes, unobserved changes in local execution quality, staffing discipline, managerial practices, or concurrent initiatives could still affect both the emergence of IRI and the impact of audits.

Future research should examine how time-varying managerial heterogeneity influences IRI formation and audit effectiveness, for example by combining operational data with survey-based or qualitative assessments of store-level routines and capabilities. Doing so would help to clarify when and how managerial attention amplifies or dampens the effects documented here.

A further limitation concerns the mechanisms linking audits, IRI corrections, and sales outcomes. The study provides evidence that correcting negative IRI is associated with higher sales, and it offers theoretically grounded arguments that this improvement likely arises from better product availability and reduced inventory freezing. However, these intermediate mechanisms are not directly observed in our data. This is a broader challenge for the research stream because IRI is not directly observable at the moment it occurs; it can only be measured ex post through deliberate audit efforts. It will therefore be difficult to develop a fuller account of the causal pathway from discrepancy emergence to stockouts and lost sales. Nevertheless, our quasi-experimental findings provide evidence consistent with the proposed mechanisms, and future research can build on this foundation by integrating inventory records with detailed on-shelf availability data, customer basket information, or observational studies of replenishment processes.

The above findings and limitations suggest two potential areas for future research. First, a better understanding of the drivers of IRI is needed. DeHoratius and Raman (2008) provided first insights into the problem. We expanded the IRI literature by identifying perishability and promotions as drivers that play an especially important role in grocery retailing. Both studies, however, use linear specifications to isolate the effect of different drivers. Future research should test for non-additive interactions among drivers, since interactions may create threshold or amplification effects overlooked by linear models. For instance, the impact of promotional activity on IRI may differ depending on an item's perishability or shelf velocity; nonlinear effects – such as threshold-based IRI increases under very high stock levels – could be explored to refine our current understanding and guide more targeted operational interventions. Similarly, both studies on the drivers of IRI are correlational, and further research is needed to identify the root causes of IRI. One promising approach in this context is machine learning that has performed well in identifying patterns that explain the variation in a variable of interest (see Chou et al., 2023, for examples and discussion of the use of machine learning for pattern identification and theory development). Only with causal explanations of IRI will we be able to quantify and address the operational issues causing it, allowing us to reduce its incidence. Developing fuller accounts of causal mechanisms, however, remains a significant challenge as, as discussed above, IRI events are not directly observable and we can only detect and report them post facto (Chuang et al., 2022). Nevertheless, further studies with broader samples and over longer periods of time should allow testing of additional potential drivers (even in a correlational sense) and reveal insights that would help develop stockcount strategies to better direct resources to those products that offer the highest sales improvement potential. Given that stockcounts are labor- and consequently cost-intensive in many sectors, retailers could contrast the cost of auditing a particular product category with the sales improvement potential of that group of SKUs. The result could be selective stockcounts with a larger audit frequency for items suffering particularly from inaccurate stock records.

The second area of future research suggested by our findings is to explore the timing of the audits for the targeted group of SKUs. This study presented some evidence of the dynamics of IRI corrections on sales with the timing effects in Models 2, 3, and 4. The results, however, are difficult to interpret as sales are also affected by other time-dependent factors (e.g., demand seasonality, holidays, promotional activity). If, as suggested by our results, the audit uplift on sales is caused by a correction of 'frozen' products – or a situation where inventory is so low that it limits sales – then, perhaps a more promising strategy would be to assess the evolution of the magnitude of IRI. Chuang et al. (2022) used empirical data to estimate the decay of stock records, from accurate to inaccurate, and developed an inspection policy that minimizes the combined cost of performing the audits and the expected cost of IRI for a group of SKUs. Their policy, however, is based on an expectation of when an item will become inaccurate and an aggregation of the error magnitudes. Our findings suggest that further precision in the description of the evolution of the magnitude of IRI has the potential for more targeted auditing policies and further cost reductions.

## Acknowledgements

The authors are grateful to the senior editor and the reviewers for their constructive comments that helped to improve an earlier version of this manuscript. The research described in this paper draws from a project that has been supported by the Efficient Consumer Response (ECR) Retail Loss Group. We are grateful to Colin Peacock from ECR for his significant role in this project, all the way from its commissioning through to ensuring the dissemination of the results. We acknowledge the various contributions of the retail community, including the valuable feedback on our research received at their events in Barcelona, Düsseldorf, Paris, Brussels, Copenhagen, and London. This paper is dedicated to the memory of Simon Parry from RGIS who passed away shortly before this paper was ready for submission. Simon was a valued member of our team and the ECR and will be sorely missed.

# Appendix

| Study | Main Objective | Context | Data Attributes (retailers, stores, SKUs) | Key findings |
|---|---|---|---|---|
| **Scope and drivers of IRI** | | | | |
| Barratt et al. (2018) | Understand how IRI develops over time and how employees interact with stock management. | Multi-channel distribution center of a retailer. | 1; 1[†]; 27 | • Substantial IRI on a daily basis with a bias towards negative errors.<br>• Channel influences IRI level.<br>• Employee-system interactions create IRI volatility. |
| Bertolini et al. (2015) | Analyze RFID effect on IRI vis-à-vis barcode technology. | Store of an Italian fashion retailer. | 1; 1; 2782 | • RFID counts are more reliable than barcode counts.<br>• Inventory is under(over) estimated by RFID(BC) counts. |
| Chuang et al. (2016) | Assess if external audits can reduce stockouts and be performed economically. | US retailer. | 1; 60; 4 | • External audits reduced shelf out-of-stocks and IRI.<br>• The audits led to higher sales. |
| Chuang and Oliva (2015) | Explore the impact of staffing levels and employees' experience and training on IRI. | Global retailer of DIY supplies. | 1; 5; ~30k | • Full-time staff reduces IRI, part-time staff does not.<br>• IRI increases workload, which can further worsen it. |
| DeHoratius and Raman (2008) | Characterize the distribution of IRI identify factors that mitigate and contribute to IRI. | Large retailer. | 1; 37; NA | • IRI are widespread and substantial.<br>• IRI is higher for fast-selling, low-cost, and low-dollar-volume items.<br>• Audit frequency and distribution method affect IRI. |
| Goyal et al. (2016) | Investigate RFID-enabled visibility as a means to decrease IRI and stockouts. | Two department store chains. | E1:1; 2; 830<br>E2:1; 2; 108 | • RFID reduced IRI ≤80%.<br>• RFID had minimal effect on backroom inventory compared to its impact on sales floor. |
| Hardgrave et al. (2009) | Investigate the impact of RFID on IRI. | Wal-Mart (air fresheners investigated). | 1; 16; 300 | • RFID reduced negative inventory discrepancies by about 13%.<br>• Manual inventory adjustments decreased with RFID implementation. |
| Hardgrave et al. (2011) | Investigate the impact of RFID on IRI of different product categories. | Large US retail chain (general merchandise). | 1; 31; 1268 | • RFID significantly reduced IRI.<br>• The impact of RFID was highest for categories with high turnover, low unit cost, and high inventory density. |
| Hardgrave et al. (2013) | Investigate the effect of improved RFID-enabled item visibility on IRI. | Global retail chain. | E1:1; 13; 337<br>E2:1; 62; 1268 | • RFID reduced IRI by ≤80%.<br>• RFID is especially effective where predictors of IRI (e.g., item variety, sales velocity) are strongest. |
| Ishfaq and Raja (2020a,b) | Descriptively evaluate IRI to educate a simulation study. | Large US apparel retailer. | 1; 1; 7634/7415 | • Inventory records contain widespread inaccuracies.<br>• Several product attributes were strongly linked to IRI profiles. |
| Rinehart (1960) | Investigate causes of IRI. | Large US federal government supply agency. | 1; 1[†]; 2,000 | • ~80% of discrepancies were caused by the very processes meant to correct them.<br>• IRI remained high after a stock audit, showing little net improvement. |
| Sheppard and Brown (1993) | Investigate factors that contribute to IRI. | Small U.S.-based electronics manufacturing company. | 1; 1[†]; 1164 | • Six key factors significantly predicted IRI.<br>• Discriminant model predicted IRI with 75% accuracy. |
| **Stock audit policies** | | | | |
| Chuang et al. (2022) | Estimates the deterioration of inventory records to propose an optimal audit policy. | Global retailer of DIY supplies. | 1; 1; 26,036 | • IRI decay can be empirically estimated using correction data.<br>• Proposed policy outperforms current store practice and industry benchmark. |
| DeHoratius et al. (2023) | Develop and test prioritization procedures for stock counting to reduce IRI cost-effectively. | DM, a European home and personal care retail company. | 1; 10; 500 | • Model-based audit prioritization performs as well as best policies across all key measures.<br>• Rule-based policies can be nearly as effective as complex models when matched to retailer goals. |
| **IRI performance impact** | | | | |
| DeHoratius and Raman (2008) | Numerical experiment to estimate sales loss caused by IRI. | Large retailer. | 1; 37; NA | • Frozen SKUs caused lost sales equal to ~1.1% of the company's annual sales and ~3.3% of its gross profit. |
| Shabani et al. (2021) | Develops measures IRI and applies them to assess store performance. | European fashion retailer based in NL. | 1; 81; 14,500 | • Proposed measure can identify stores where performance inefficiency is linked to IRI. |
| Ton and Raman (2010) | Examines drivers of phantom products and their impact on sales. | Borders Group Inc., a book and stationary retailer. | 1; 356; NA | • Higher inventory raise phantom products, which indirectly lower sales.<br>• More inventory and variety increase sales, but also increase operational complexity, causing negative effects through phantom products. |

[†] The authors did not investigate a retail store, but a distribution center, supply facility or warehouse.